\documentclass[aps,prd,twocolumn,eqsecnum]{revtex4-1}
\usepackage{hyperref}
\usepackage{graphicx}
\usepackage{amsmath}
\usepackage[utf8]{inputenc}

\begin{document}

\title{Spherical ansatz for parameter-space metrics}

\date{\today}

\author{Bruce Allen}
\affiliation{MPI for Gravitational Physics, Callinstrasse 38, Hannover, Germany}

\begin{abstract}
  \noindent
  A fundamental quantity in signal analysis is the metric $g_{ab}$ on
  parameter space, which quantifies the fractional ``mismatch'' $m$
  between two (time- or frequency-domain) waveforms.  When searching
  for weak gravitational-wave or electromagnetic signals from sources
  with unknown parameters $\lambda$ (masses, sky locations,
  frequencies, etc.) the metric can be used to create and/or
  characterize ``template banks''. These are grids of points in
  parameter space; the metric is used to ensure that the points are
  correctly separated from one another. For small coordinate
  separations $d\lambda^a$ between two points in parameter space, the
  traditional ansatz for the mismatch is a quadratic form $m=g_{ab}
  d\lambda^a d\lambda^b$. This is a good approximation for small
  separations but at large separations it diverges, whereas the actual
  mismatch is bounded.  Here we introduce and discuss a simple
  ``spherical'' ansatz for the mismatch $m=\sin^2(\sqrt{g_{ab}
    d\lambda^a d\lambda^b})$. This agrees with the metric ansatz for
  small separations, but we show that in simple cases it provides a
  better (and bounded) approximation for large separations, and argue
  that this is also true in the generic case.  This ansatz should
  provide a more accurate approximation of the mismatch for
  semi-coherent searches, and may also be of use when creating grids
  for hierarchical searches that (in some stages) operate at
  relatively large mismatch.
\end{abstract}


\maketitle

\section{Matched filtering and the overlap between templates}
\label{s:intro}

More than two decades ago, when the first generation of
interferometric gravitational wave (GW) detectors were still in the
planning stages, a handful of pioneers investigated the techniques
that would be needed to detect GW signals \cite{1989CQGra...6.1761S,
  1989ASIC..253.....S, 1991PhRvD..44.3819S, 1991dgw..conf..406S,
  1993PhRvD..47.2198F, 1993PhRvL..70.2984C, 1994PhRvD..50.7111S,
  1994PhRvD..49.2658C, 1994PhRvD..49.6274A, 1995PhRvD..52..605A}.  At
that time there were three main challenges.  First, the signals were
weak in comparison with the noise from the detectors, so needed to be
``teased out'' of the data stream with optimal or near-optimal
methods.  Second, the parameters describing the signals (such as the
object masses in a binary system, or the rotation frequency and
spindown rate of a neutron star) were not known. This required repeated
searches for signals with many different parameter combinations,
creating a significant computational challenge.  Lastly, even if the
parameters were known precisely, for some sources the waveforms could
only be calculated approximately.  The errors could be estimated but
not sharply quantified.

The solution to the first problem is to use ``matched filtering''
\cite{1991PhRvD..44.3819S, 1993PhRvL..70.2984C, 1994PhRvD..49.1707D,
  1994PhRvD..50.2390D, 1994PhRvD..49.2658C, 1994PhRvD..50.6080B,
  1994PhRvD..50.7111S, 1996PhRvD..53.3033B, 1996PhRvD..54R1860B,
  1999PhRvD..60b2002O}.  In the simplest case \footnote{In many cases
  of interest a quantity corresponding to SNR is analytically
  maximized over some intrinsic phase parameters.  This results in a
  ``squared-SNR'' detection statistic which is quadratic in the data.
  The mismatch we define then corresponds to the fractional loss of
  expected squared SNR in the strong signal limit.} the time-dependent
output $S(t)$ of the detector is correlated with a template $T(t)$ to
produce a statistic
\begin{equation}
  \rho = (T , S).
  \label{e:rho}
\end{equation}
If the template is normalized $(T,T)=1$ then $\rho$ is called the
signal-to-noise ratio (SNR). This is reviewed in a signal-processing
context in \cite{ 1970esn..book.....W, 1970PhT....23f..73H} and in the
GW context in \cite{2009agwd.book.....J} and
\cite{2011gwpa.book.....C}.

The positive-definite inner product in Eq.~(\ref{e:rho}) can be
expressed in different ways.  For example if the instrument noise is
white (or the signal is confined to a narrow enough range of frequency
that the noise is white in that band) then the inner product is
\begin{equation}
 (A , B) = {\cal N} \int A(t) B(t) \; dt,
  \label{e:innerprod_time}
\end{equation}
where the integral extends over the support of the waveform or the
duration of the data (whichever is shorter).  The normalization
constant $\cal{N}$ is set by requiring that the expected value of
$(S,S)$ is unity where $S(t)$ is the detector output in the absence of
any signals \footnote{If the signal is narrow-band, then to get the
  most SNR the instrument output should be filtered so that $n(t)$
  only contains the band of interest}.

If the detector noise is colored \cite{1994PhRvD..49.1707D} then the
inner product is most simply expressed in the frequency domain as
\begin{equation}
 (A , B) = \int_{-\infty}^{\; \infty}  \frac{{\tilde A}^* (f) {\tilde B}(f)}{S(|f|)} \; df.
  \label{e:innerprod_freq}
\end{equation}
Here, the Fourier transform of a function of time $h(t)$ is denoted by
$\tilde h(f)$, where $f$ is frequency, and $S(f)$ is the (single-sided)
noise power spectrum of the instrument.

For real instrument data sampled at a finite rate the integral in
Eq.~(\ref{e:innerprod_time}) may be replaced with a sum over samples and
the integral in Eq.~(\ref{e:innerprod_freq}) may be replaced with a sum
over Nyquist-sampled frequency bins \cite{2012PhRvD..85l2006A}.

The solution to the second problem is to construct the SNR $\rho$ in
Eq.~(\ref{e:rho}) for many different templates $T_{\lambda_i}$, where
$\lambda$ are the parameters that describe the waveform and the
integer $i$ labels a finite set of distinct points which are being
sampled from parameter space \cite{1991PhRvD..44.3819S,
  1994PhRvD..49.1707D, 1995PhRvD..52..605A, 1996PhRvD..54.2421A,
  1999PhRvD..60b2002O, 2004CQGra..21S1635C, 2006CQGra..23.5477B,
  2009PhRvD..80j4014H, 2009PhRvD..79j4017M,
  2010PhRvD..81b4004M,2013PhRvD..87h2004B, 2014PhRvD..90l4049F,
  2014PhRvD..89d2002K, 2017PhRvD..95j4045R}.  $\lambda$ denotes the
collection of coordinates in parameter space; the individual
coordinates are denoted by $\lambda^a$ where the index $a = 1, 2,
\cdots, N$ runs over the parameter-space coordinates.

For GWs from compact binary coalescence (CBC) $\lambda$ includes the
masses of the objects, sky location, orbital inclination, time of the
merger, spins (if relevant) and so on.  For continuous gravitational
waves (CW) from a spinning neutron star $\lambda$ includes the sky
location, frequency and frequency derivative, and so on \footnote{Some
  parameters are NOT included in $\lambda$.  For example the
  (normalized) templates are independent of the distance to the
  source.  For search techniques based on fast Fourier transforms
  (FFTs) the coalescence time or signal phase might be left out
  because they are automatically searched over.}
\cite{1998PhRvD..57.2101B, 2000PhRvD..61h2001B, 1998PhRvD..58f3001J,
  2011PhRvD..83d3001P, 2018PhRvD..97l3016W}.

The set of templates is called a template bank, and the art is in
selecting their locations $\lambda_i$ \cite{1996PhRvD..53.6749O,
  1999PhRvD..60b2002O, 2003PhDT.........3C, 2004CQGra..21S1635C,
  2004PhRvD..70l2001C, 2006CQGra..23.5477B, 2006IJMPD..15..225S,
  2006PhDT........24P, 2007CQGra..24.6227C, 2007CQGra..24S.481P,
  2007PhRvD..76j2004C, 2008CQGra..25r4029W, 2008CQGra..25s5011B,
  2008CQGra..25s5011B, 2008PhRvD..77j4017A, 2009CQGra..26d5013C,
  2009PhRvD..79j4017M, 2009PhRvD..79l9901A, 2009PhRvD..80b4009V,
  2009PhRvD..80j4014H, 2010CQGra..27e5010W, 2010PhRvD..81b4004M,
  2011PhRvD..83d3001P, 2013PhRvD..87h2004B, 2013PhRvD..87l4003K,
  2014PhRvD..89d2002K, 2014PhRvD..89h4041A, 2014PhRvD..90l4049F,
  2015CQGra..32n5014P, 2016PhRvD..93l2001C, 2017PhRvD..95f4056I,
  2017PhRvD..95j4045R, 2017arXiv170501845D, 2017arXiv171108743R,
  2018PhRvD..97l3016W, 2018arXiv181205121M, 2018cosp...42E2899R,
  2019PhRvD..99b4048R, 2019arXiv190401683R}.  Since the signals
themselves come from a continuous family and the template bank is
discrete, real signals will not have parameters that exactly match any
template in the bank.  So one must ensure that there is at least one
template ``close enough'' to the signal that it is not missed.  At the
same time, since $\rho$ must be computed for each template, the number
of templates should be no larger than needed.  For the advanced LIGO
and Virgo instruments, the CBC searches employ $O(10^5)$ templates;
the CW searches employ orders of magnitude more.

To place the templates in parameter space, an important quantity is
the overlap (also called a fitting function or match) between two
templates
\begin{equation}
  o(\lambda, \lambda') = (T_\lambda,T_{\lambda'}).
  \label{e:overlap1}
\end{equation}
Because the templates are normalized, and the inner product is
positive definite, the overlap lies in the closed interval $o \in
[-1,1]$ \footnote{Typically waveforms can have either sign, so one can
  constrain this overlap to the closed interval $[0,1]$.}.

The overlap is also relevant to the third of the challenges described
above because it may be used to quantify the loss in SNR arising from
inaccuracies in the waveform models.  However in this paper we assume
that the waveform models are exact, and concentrate on the previous
issue.

\section{The mismatch and the metric approximation to the mismatch}
\label{s:mismatchand metric}

Rather than using the overlap, it is more convenient to use a related
quantity called the mismatch, but the literature contains several
different definitions for this.  Much of the work on CBC data analysis
uses the mismatch $1-o$ and much of the literature on CW signals uses
$1-o^2$.  Here, we follow the latter convention, defining the mismatch
as
\begin{equation}
  m(\lambda, \lambda') = 1 - o^2(\lambda, \lambda').
  \label{e:truemismatch}
\end{equation}
This mismatch lies in the interval $[0,1]$ and is the fractional loss
in the square of the expected SNR $\langle \rho \rangle ^2$ that
arises when a signal with parameters $\lambda$ is detected using a
template with parameters $\lambda'$.  Sec.~\ref{s:twodefs} gives
results for another common definition, where the mismatch is the
fractional loss of the expected SNR $\langle \rho \rangle $.  In the
Neyman–Pearson approach, $m$ is the fractional loss in the maximum of
the log likelihood ratio in the strong signal limit.

We note that a signal search algorithm may (either analytically or
explicitly) minimize the mismatch with respect to some of the
intrinsic or extrinsic parameters.  In this case we assume that these
parameters are not included in the vector $\lambda$ and that the
right-hand side (rhs) of the expression in Eq.~(\ref{e:truemismatch})
for $m$ is minimized over those missing parameters
\footnote{ Suppose that the mismatch is minimized with respect to
  template parameters $\alpha, \cdots, \beta$, and that the templates
  are a continuous function of these parameters. Then the results
  given here still hold, because at the extrema one may express each
  of these parameters as a function of the remaining free parameters,
  i.e. one has $\alpha(\lambda, \lambda'),\cdots, \beta(\lambda,
  \lambda')$.} In hierarchical searches, the mismatch may also be
averaged over data segments; we return to this in Sec.~\ref{s:sowhat}.

\begin{figure}[htbp]
    \includegraphics[width=0.5\textwidth]{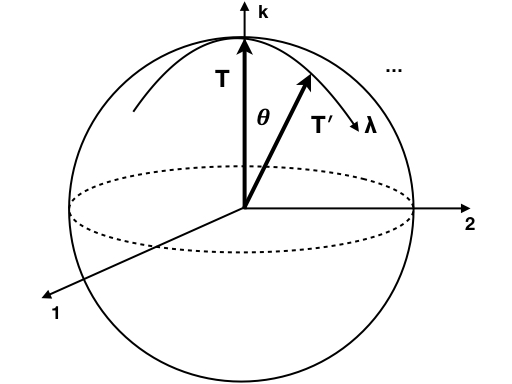}
    \caption{\label{f:sphere} The normalized templates $T=T(\lambda)$
      and $T'=T(\lambda')$ may be thought of as unit vectors lying on
      the surface of a $(k-1)$-sphere, where the embedding dimension
      $k$ is the number of discrete time-domain samples in the
      waveform (only three of these dimensions are shown here).  A
      one-dimensional variation of the parameters $\lambda$ traces out
      a path on the sphere, where the angular separation $\theta$
      between the points is defined by $\cos \theta = (T, T')$.}
\end{figure}

It is helpful to think of the normalized templates $T=T_\lambda$ and
$T'=T_{\lambda'}$ as unit-length vectors which lie on the surface of
the unit sphere $S^{k-1}$ as illustrated in Fig.~\ref{f:sphere}.  In
the case where the data and template are discretely sampled, $k$ is
the number of samples in the template.  In the continuous case $k$ is
infinite and the sphere is embedded in a Hilbert space \footnote{This
  is very similar to the way that a normalized wave-function in
  quantum mechanics is expressed as an infinite sum of coefficients
  multiplying basis vectors.}.

We define the angle $\theta(\lambda, \lambda')$ between two normalized
templates via
\begin{equation}
\cos \theta(\lambda, \lambda') = o(\lambda,\lambda')= (T, T')
\end{equation}
so that the mismatch may be expressed as
\begin{equation}
  m(\lambda, \lambda') = 1 - \cos^2 \theta = \sin^2 \theta.
\end{equation}
Since the mismatch is extremal and vanishes at $\lambda=\lambda'$ it
can be expanded in a Taylor series which (generically) begins at
quadratic order.

This ``metric approximation'' to the mismatch has a geometrical
interpretation which was introduced in \cite{1996PhRvD..53.6749O} and
elaborated in \cite{1996PhRvD..53.3033B,1996PhRvD..54R1860B,
  1999PhRvD..60b2002O}.  It is
\begin{equation}
  m(\lambda, \lambda') = g_{ab} d\lambda^a d\lambda^b + O(q_{abc} d\lambda^a d\lambda^b d\lambda^c ),
\end{equation}
where $d\lambda=\lambda -\lambda'$, and we adopt the ``Einstein
summation convention'' that repeated indices $a, b, \cdots , c$ are
summed from $1$ to $N$.  The quantity $g_{ab}$ is called the
parameter-space metric \cite{1996PhRvD..53.6749O}; for nearby
templates, $g_{ab} d\lambda^a d\lambda^b$ measures the squared
fractional deviation or squared dimensionless ``interval'' between the
templates.

We note that there are other possible definitions of the metric, but
this choice is normally adopted for the template placement problem,
because templates must be placed ``independently of the data'' based
on the expected properties of the signals and detector noise.  A good
discussion of this and of other possible definitions of
parameter-space metrics may be found in the Introduction and in
Appendix A of Prix \cite{2007PhRvD..75b3004P,2007PhRvD..75f9901P}.

\section{Simple illustrative example}
\label{s:simpleexample}

To make this concrete, we consider a simple CW example.  The waveforms
are described by a single (angular) frequency parameter $\lambda^1 =
\omega$.  In the time domain the normalized templates are
\begin{equation}
  T_\omega(t) = \sqrt{\frac{1}{\tau}} \sin( \omega t ) \quad {\rm for} \quad t \in [-\tau,\tau]
\end{equation}
and vanish for $|t|>\tau$.  In the cases of interest $\tau$ would be
days to years, and $\omega$ would be tens to hundreds of cycles per
second.

The overlap and mismatch between two templates may be easily computed,
starting from Eq.~(\ref{e:innerprod_time}) with ${\cal N} = 1$:
\begin{eqnarray}
  \nonumber
  (T, T') & = & \frac{1}{\tau} \int_{-\tau}^{\tau} \sin(\omega t) \sin(\omega' t) dt \\
  & = & \frac{ \sin (\omega - \omega') \tau } {(\omega - \omega') \tau} - \frac{ \sin (\omega + \omega') \tau } {(\omega + \omega') \tau}.
  \label{e:sinc}
\end{eqnarray}
For the cases of interest $\omega$ is large enough that there are many
cycles in the interval $t \in [-\tau, \tau]$, and the fractional
difference between $\omega$ and $\omega'$ is small.  This means that
the second term on the rhs of Eq.~(\ref{e:sinc}) is
negligible, so the mismatch is given by the square of the sinc
function
\begin{equation}
  m = 1-(T,T')^2 = 1 - \biggl[ \frac{ \sin \tau \Delta \omega } {\tau \Delta \omega} \biggr]^2,
  \label{e:examplemismatch}
\end{equation}
where $\Delta \omega= \omega-\omega'$.  This may be expanded as a
Taylor series for small $\Delta \omega$, yielding $ m = \frac{1}{3}
\tau^2 \Delta \omega^2 + O(\Delta \omega^4)$.  Thus the metric is
$g_{\omega \omega} = \tau^2/3$ and the metric approximation to the
mismatch is
\begin{equation}
  m = \frac{1}{3} \tau^2 \Delta \omega^2.
  \label{e:examplemetric}
\end{equation}

\section{The metric approximation and the spherical approximation}
\label{s:metricversusspherical}

\begin{figure}[htbp]
    \includegraphics[width=0.5\textwidth]{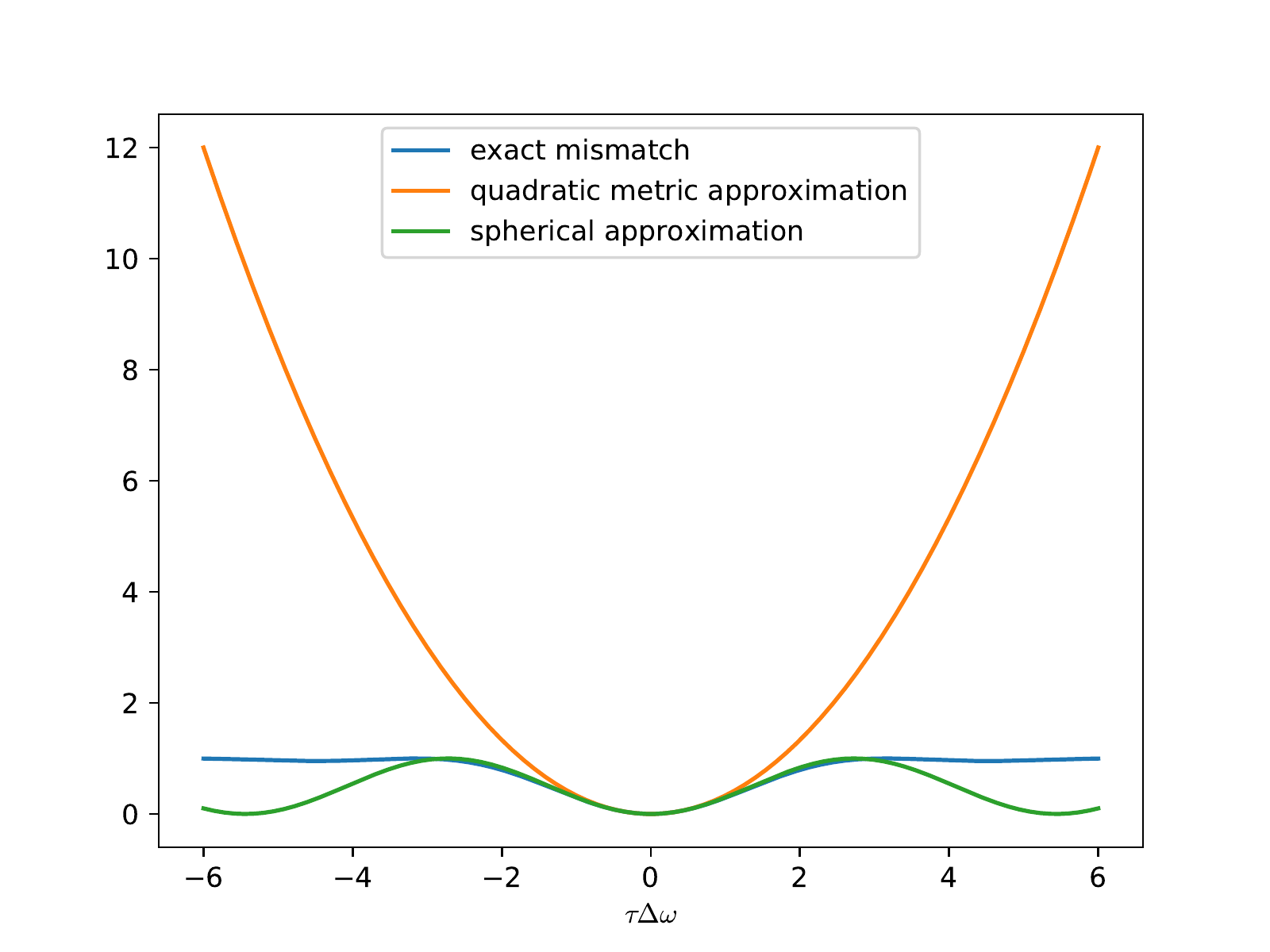}
    \caption{\label{f:plots} The blue curve shows the mismatch $m$
      from Eq.~(\ref{e:examplemismatch}) for long-duration sinusoidal
      signals of frequency $\omega$ and $\omega'$, as a function of
      the frequency difference $\Delta \omega = \omega - \omega'$.
      The orange curve is the conventional metric approximation to the
      mismatch, given in Eq.~(\ref{e:examplemetric}).  The green curve
      shows the spherical approximation to the mismatch, given in
      Eq.~(\ref{e:examplespherical}).  While both approximations agree
      for small $|\Delta \omega|$, one can see that the spherical
      approximation is more accurate and has a larger domain of
      applicability. This is generically true because the spherical
      ansatz has one approximation fewer than the conventional metric
      ansatz.}
\end{figure}

Shown in Fig.~\ref{f:plots} (blue) is the actual mismatch $m$ as a
function of $\lambda-\lambda' = \Delta \omega$, as given by
Eq.~(\ref{e:examplemismatch}).  Also shown (orange) is the metric
approximation from Eq.~(\ref{e:examplemetric}).  One can see that these
agree well for small values of $\Delta \omega$, but that the metric
approximation breaks down when $| \Delta \omega | \gtrsim 1/\tau$.
One can also see that where they deviate, the quadratic approximation
tends to {\it overestimate} the mismatch.  This is well known to the
experts \footnote{About two decades ago, Benjamin Owen pointed this
  out to me, and told me that it was an effect arising from
  ``projection onto the sphere''. He was right!}  and frequently
observed when the metric approximation is compared to the true
mismatch.  Below we provide both the explanation and a simple
solution.

It is helpful to visualize this on the sphere.  Imagine that we have a
path in parameter space, parameterized by the variable $\lambda$ as
shown in Fig.~\ref{f:sphere}, which passes through the template $T$ at
parameter value $\lambda$, and $T'$ at parameter value $\lambda'$.  A
generic parameterization is one for which the angle $\theta$ varies
linearly with $\lambda-\lambda'$ for small values of
$\theta$ \footnote{Note that the exact behavior of the mismatch as a
  function of the parameter coordinate separation depends upon the
  parameterization of the waveforms.  Any non-linear transformation of
  the parameter will change this behavior, but a ``generic'' choice of
  parameterization will not. A generic parameterization is one for
  which the angle $\theta$ sweeps steadily along the sphere.}.

For such a generic parameterization, the angular separation on the
sphere is well approximated by
\begin{equation}
  \theta = \sqrt{g_{ab} d\lambda^a d\lambda^b}.
\label{e:thetadef}
\end{equation}
This means that the mismatch can be written (in what we here call the
``spherical approximation'') as
\begin{equation}
  m = \sin^2 \theta = \sin^2 \sqrt{g_{ab} d\lambda^a d\lambda^b}.
  \label{e:betterapprox}
\end{equation}
The point of this short paper is that Eq.~(\ref{e:betterapprox}) is a
better approximation to the generic mismatch than the more
conventional approximation $m=g_{ab} d\lambda^a d\lambda^b$.  While
both of these approximations agree to lowest order in the parameter
separation $d\lambda^a$, for the generic case the approximation given
in Eq.~(\ref{e:betterapprox}) will be accurate for a larger range in
$d\lambda^a$.  It also has the advantage of always lying in the
interval $[0,1]$.

The simple example presented in Sec.~\ref{s:simpleexample} is a good
demonstration of this.  Fig.~\ref{f:plots} shows how the behavior of
the conventional metric approximation (orange curve) deviates from the
actual mismatch (blue curve) as the parameter mismatch $\Delta \omega$
increases.  The spherical approximation (green curve) given by
\begin{equation}
  m = \sin^2 \sqrt{g_{ab} d\lambda^a d\lambda^b} = \sin^2 \frac{\tau \Delta \omega}{\sqrt{3}}
  \label{e:examplespherical}
\end{equation}
is a much better fit to the actual mismatch.  It deviates
significantly from the actual mismatch only after the path on the
sphere has exceeded a 90-degree separation between $T$ and $T'$.

\section{A second example}

The reader might ask if we have ``tuned'' our example in
Sec.~\ref{s:simpleexample}.  This is not so: the dependence upon the
parameter (frequency) is quite typical. Here we present another
typical case, where the signal model depends upon an offset phase
parameter $\phi$.

For this example, the normalized time domain templates are
\begin{equation}
  T_\phi(t) = \sqrt{\frac{1}{\tau}} \sin( \omega t + \phi ) \quad {\rm
    for} \quad t \in [-\tau,\tau],
\end{equation}
and vanish for $|t|>\tau$. As in our previous example, we assume that
the signal goes through many cycles in the observation interval, so
that $\omega \tau \gg 1$ is large.

The overlap between templates is easily calculated from
\ref{e:innerprod_time}, giving
\begin{eqnarray}
  \nonumber
  (T, T') & = & (T_\phi, T_{\phi'}) \\
  \nonumber
  & = &  \frac{1}{\tau} \int_{-\tau}^{\tau} \sin(\omega t + \phi) \sin(\omega t + \phi') dt \\
  & = & \cos (\phi - \phi') - \frac{ \sin  2 \omega \tau }{2\omega \tau} \cos ( \phi + \phi' ).
  \label{e:phioverlap}
\end{eqnarray}
Since we are assuming that $\omega \tau \gg 1$, the second term on the
rhs can be neglected, giving the mismatch
\begin{equation}
    m = 1-(T,T')^2 = \sin^2 \Delta \phi,
\end{equation}
where $\Delta \phi = \phi - \phi'$.

In this case, the metric approximation to the mismatch yields the
quadratic form $m = g_{ab} d\lambda^a d\lambda^b = (\Delta \phi)^2$.
In contrast, the spherical approximation to the mismatch gives
\begin{equation}
    m = \sin^2 \sqrt{g_{ab} \lambda^a d\lambda^b } = \sin^2 \Delta \phi.
\end{equation}
So in our second example, the spherical approximation is exact!

\section{A third example}

In our final example, the signal parameter is a constant frequency
derivative $\lambda^1 = \dot \omega$. The normalized time domain
templates are
\begin{equation}
  T_{\dot \omega}(t) = \sqrt{\frac{1}{\tau}} \sin( \omega t + \dot \omega t^2/2 ) \quad {\rm
    for} \quad t \in [-\tau,\tau],
\end{equation}
and vanish for $|t|>\tau$.  We assume that (half of the) dimensionless
phase accumulated during the observation time $\omega \tau + \dot
\omega \tau^2/2$ is much larger than $2 \pi$.

The overlap between templates is
\begin{eqnarray}
  \nonumber
  (T, T') & = & (T_{\dot \omega}, T_{\dot \omega'}) \\
  \nonumber
  & = &  \frac{1}{\tau} \int_{-\tau}^{\tau} \sin({\omega t + \dot \omega} t^2/2 )  \sin({\omega t + \dot \omega}' t^2/2 )\; dt \\
  \nonumber
  & = &  \frac{1}{2 \tau} \int_{-\tau}^{\tau} \cos(|\Delta \dot \omega |  t^2/2)\; dt \\
  & = &  \sqrt{\frac{\pi}{ | \Delta \dot \omega | \tau^2 }} C\biggl( \sqrt{\frac{ | \Delta \dot \omega | \tau^2 }{\pi}} \biggr),
  \label{e:fdotoverlap}
\end{eqnarray}
where $C(z)$ is the Fresnel integral function, $\Delta \dot \omega =
\dot \omega - \dot \omega' $, and we have dropped small terms from the
rhs in the third line of Eq.~(\ref{e:fdotoverlap}).

\begin{figure}[htbp]
    \includegraphics[width=0.5\textwidth]{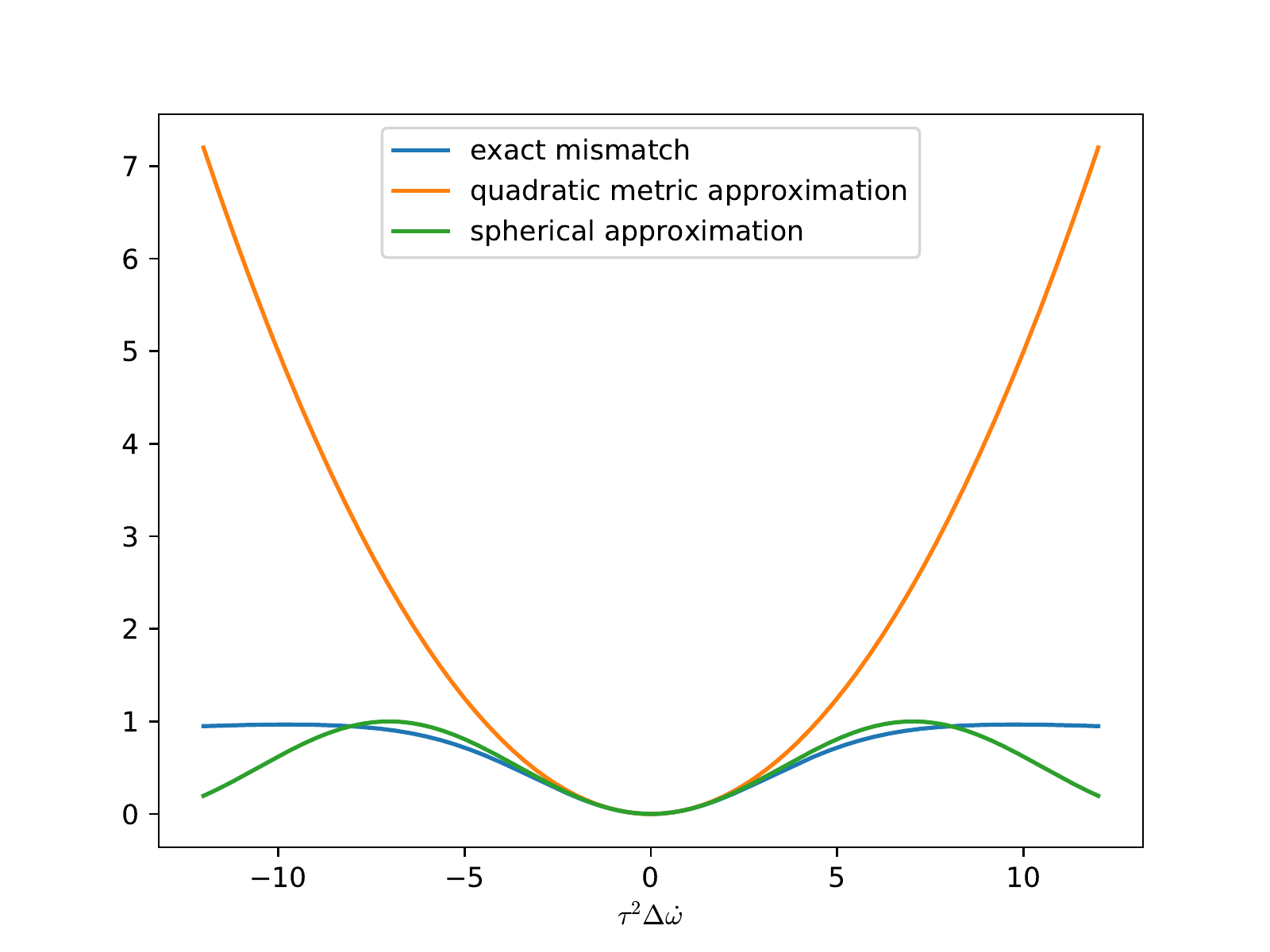}
    \caption{\label{f:plots3} The blue curve shows the true mismatch
      $m$ from Eq.~(\ref{e:exactmissdotomega}) for long-duration sinusoidal
      signals with constant frequency derivative, as a function of the
      frequency derivative difference $\Delta \dot \omega = \dot
      \omega - \dot \omega'$.  The orange curve is the conventional
      metric approximation to the mismatch, and the green curve shows
      the spherical approximation to the mismatch. As in the other
      examples, the spherical approximation is more accurate.}
\end{figure}

The exact mismatch is given by
\begin{equation}
  m(\dot \omega, \dot \omega') = 1 - \frac{\pi}{ | \Delta \dot \omega | \tau^2} C^2\biggl( \sqrt{\frac{ | \Delta \dot \omega | \tau^2}{\pi}} \biggr).
  \label{e:exactmissdotomega}
\end{equation}
This is plotted in blue in Fig.~~\ref{f:plots3}.  Since $C(z) = z -
\frac{\pi^2}{40} z^5 + O(z^9)$ for small z, the normal metric
approximation to the mismatch is $m = \tau^4 \Delta \dot \omega^2 /
20$.  This is plotted in orange.  Shown in green is the spherical
approximation to the mismatch, $m = \sin^2(\tau^2 \Delta \dot
\omega/\sqrt{20})$.

As in the previous examples, the spherical approximation is a better
fit to the true mismatch.

\section{Why does it matter?}
\label{s:sowhat}

Why does this matter? After all, the metric approximation is only
defined to quadratic order, and the mismatch can be expanded to higher
order if needed. The point here is that there are really two
approximations taking place.  The first is in the Taylor approximation
of the separation $\theta$ on the sphere, and the second is in the
Taylor approximation of the sin function in the expression $\sin^2
\theta$ which relates the mismatch to $\theta$.  The conventional
metric ansatz makes {\it both} of these approximations, whereas the
spherical ansatz only uses the first of these approximations. So for
generic behavior of the path in parameter space, the spherical ansatz
will be more accurate than the metric ansatz.  And since the spherical
approximation just replaces $m=g_{ab}d\lambda^a d\lambda^b$ with
$m=\sin^2 \sqrt{g_{ab}d\lambda^a d\lambda^b}$, this comes with no
additional analytic or computational cost.

A more accurate approximation is useful because the metric is often
used to construct grids in parameter space \cite{ 1996PhRvD..54.7108M,
  1998PhRvD..57..630M, 1999PhRvL..83.1498A, 2004CQGra..21S1635C,
  2006CQGra..23.5477B, 2007CQGra..24S.481P, 2007CQGra..24S.595B,
  2007PhRvD..76j2004C, 2008PhRvD..77j4017A, 2008CQGra..25s5011B,
  2009PhRvD..79j4017M, 2009PhRvD..80j4014H, 2010PhRvD..81b4004M,
  2011PhRvD..83d3001P, 2013PhRvD..87h2004B, 2014PhRvD..89d2002K,
  2014PhRvD..90l4049F, 2015PhRvD..92h2003W, 2017PhRvD..95j4045R}.  In
situations where a search is not compute-power limited, these grids
typically have a low mismatch.  For example in CBC searches the
traditional SNR mismatch is chosen at 3\%, corresponding to an SNR$^2$
mismatch of 6\%.  For such small mismatches, the fact that $\sin
\theta \approx \theta$ for small $\theta$ means that there is no
significant difference between the metric and spherical
approximations.  However this may not be so for searches which are
compute-power limited, for example in the search for CWs or the search
for gamma-ray pulsars.

These computationally-limited searches often employ multiple
hierarchical stages, which mix semi-coherent and coherent stages, each
employing its own metric for template placement \cite{
  2007PhRvD..75b3004P, 2007PhRvD..75f9901P, 2007CQGra..24S.481P,
  2009PhRvD..79j4017M, 2009PhRvL.103r1102P, 2010PhRvD..82d2002P,
  2011PhRvD..83l2003P, 2012ApJ...744..105P, 2013PhRvD..87h4057S,
  2013PhRvD..88l3005W, 2014ApJ...795...75P, 2015PhRvD..91j2003L,
  2017PhRvD..96f2002A, 2017PhRvD..96l2004A, 2017ApJ...834..106C,
  2018PhRvD..97l3016W, 2018PhRvD..98h4058D,
  2018IAUS..337..382N}. Those hierarchical stages sometimes operate at
substantial mismatches in the range $m \in [0.5, 0.7]$, and here, the
spherical approximation is an improvement on the conventional
quadratic approximation.

We can illustrate this using the example from
Sec.~\ref{s:simpleexample}.  Suppose we set up a one-dimensional grid
in frequency $\omega$, with a spacing $\Delta \omega$ picked to give a
desired mismatch $m$.  The metric approximation in
Eq.~(\ref{e:examplemetric}) gives a parameter-space grid spacing
\begin{equation}
  \Delta \omega = \frac{\sqrt{3}}{\tau} \; \sqrt{m},
\end{equation}
whereas the spherical approximation gives a grid spacing
\begin{equation}
  \Delta \omega = \frac{\sqrt{3}}{\tau} \; \arcsin \sqrt{m} .
\end{equation}
Effectively, the spherical approximation amounts to replacing the
conventional metric mismatch $m$ with $(\arcsin \sqrt{m})^2$.  The
effect of this on the grid spacings is shown in
Table~\ref{t:comparison}.

\begin{table}
  \begin{tabular}{c | c | c}
    Mismatch &   Metric approximation       & Spherical approximation \\
    $m$      & grid spacing $\Delta \omega$ & grid spacing  $\Delta \omega$ \\
    \hline
    0.01 & 0.173/$\tau$ & 0.173/$\tau$ \\
    0.02 & 0.245/$\tau$ & 0.246/$\tau$ \\
    0.05 & 0.387/$\tau$ & 0.391/$\tau$ \\
    \hline
    0.1 & 0.548/$\tau$  & 0.557/$\tau$ \\
    0.2 & 0.775/$\tau$  & 0.803/$\tau$ \\
    0.5 & 1.225/$\tau$  & 1.360/$\tau$ \\
    \hline
    0.7 & 1.449/$\tau$ & 1.717/$\tau$ \\
    0.9 & 1.643/$\tau$ & 2.163/$\tau$ \\
  \end{tabular}
  \caption{\label{t:comparison} One-dimensional grid spacings $\Delta
    \omega$ for the simple example in Sec.~\ref{s:simpleexample},
    comparing conventional versus spherical approximation to the
    metric.  These agree for small mismatch, but diverge for
    mismatches approaching unity. Such large mismatches may be used in
    (multi-stage hierarchical) searches which are computing-power
    limited.}
\end{table}

The spherical approximation might also provide a significant
improvement for semi-coherent searches, when compared with the normal
quadratic metric approximation.  Semi-coherent methods are employed
for computationally-limited electromagnetic and GW searches, and
consist of breaking a long data stream into $M$ shorter
``computationally-feasible'' segments, each of which is searched using
traditional matched-filter methods.  The resulting ``coherent''
statistics (typically SNR values) are then summed to produce the
semi-coherent statistic, as first proposed in
\cite{1998PhRvD..57.2101B, 2000PhRvD..61h2001B}.  To set up a template
grid one computes a semi-coherent metric $\bar g_{ab}$ to predict the
fractional loss of the semi-coherent statistic. Until now $\bar
g_{ab}$ has been computed by summing or averaging the coherent
metrics $g_{ab}^{(i)}$ for the $i=1, \cdots, M$ segments of the
coherent searches \cite{2009PhRvL.103r1102P, 2010PhRvD..82d2002P,
  2013PhRvD..88l3005W, 2015PhRvD..92h2003W,2015PhRvD..91j2003L}.  This
averaged metric can be a poor approximation, and recent work has
investigated its accuracy and empirical ways to extend the range of
validity \cite{2016PhRvD..94l2002W}.

This work suggests a possible improvement.  Instead of estimating the
semi-coherent mismatch with an averaged metric
\begin{equation}
  m = \bar g_{ab} d\lambda^a d\lambda^b  = \frac{1}{M} \sum_{i=1}^{M} g_{ab}^{(i)}  d\lambda^a d\lambda^b ,
  \label{e:approx1}
\end{equation}
it might be more accurate to instead compute the semi-coherent
mismatch in the spherical approximation:
\begin{equation}
  m = \frac{1}{M} \sum_{i=1}^{M} \sin^2\sqrt{g_{ab}^{(i)} d\lambda^a d\lambda^b }.
  \label{e:approx2}
\end{equation}
Because the sin-squared of the average is not the average of the
sin-squared, Eqs.~\ref{e:approx1} and \ref{e:approx2} could differ
substantially, particularly if the quadratic approximation to the
metric significantly overestimates the mismatch in one or more of the
coherent segments.

\section{Conventions for overlap and mismatch}
\label{s:twodefs}

In much of the CBC literature the mismatch is defined as
\begin{equation}
  m(\lambda, \lambda') = 1-o(\lambda, \lambda').
\end{equation}
This SNR fractional mismatch should be contrasted with the SNR${}^2$
fractional mismatch defined in Eq.~(\ref{e:truemismatch}).  With this
definition of the mismatch, the same considerations as above give the
spherical approximation as
\begin{eqnarray*}
  m & = & 1 - \cos \sqrt{g_{ab} d\lambda^a d\lambda^b} \\
    & = & 2 \sin^2  \frac{1}{2} \sqrt{g_{ab} d\lambda^a d\lambda^b}.
\end{eqnarray*}
This should be contrasted with the spherical approximation given in
Eq.~(\ref{e:betterapprox}).

\section{Conclusion}

For three typical examples, we have shown that replacing the
conventional metric mismatch $g_{ab} d\lambda^a d\lambda^b$ with
$\sin^2 \sqrt{g_{ab} d\lambda^a d\lambda^b}$ gives a better
approximation to the true template mismatch. We have argued that this
is to be expected in the generic case, and suggested that averaging
the spherical approximation might provide a more accurate way to
compute the mismatch in semi-coherent searches.

I thank Curt Cutler, Maria Alessandra Papa and Reinhard Prix for
encouragement and helpful comments, and Ben Owen and
B.S. Sathyaprakash for teaching me about matched filtering and
metrics.

\newpage
\bibliography{references}

\end{document}